\documentclass[preprint, 12pt]{aastex}

\begin{document}

\title{A Search for Near-Infrared Emission From the Halo of NGC 5907 at
Radii of 10 kpc to 30 kpc}

\author{
Sarah A. Yost\altaffilmark{1}\altaffiltext{1}{California Institute of
Technology, Pasadena, CA 91125}, James J. Bock\altaffilmark{2}
\altaffiltext{2}{Jet Propulsion Laboratory, California Institute of
Technology, 4800 Oak Grove Drive, Pasadena, CA 91109}, 
Mitsunobu Kawada\altaffilmark{4} \altaffiltext{3}{Institute of Space and
Astronautical Science, 3-1-1, Yoshinodai, Sagamihara-shi, Kanagawa
229-8510, Japan}, Andrew E. Lange\altaffilmark{1},
Toshio Matsumoto\altaffilmark{3}, Kazunori Uemizu\altaffilmark{3}, Toyoki
Watabe\altaffilmark{4} \altaffiltext{4}{Nagoya University, Chikusa-ku,
Nagoya 464-8602, Japan}, \and Takehiko Wada\altaffilmark{3}}

\begin{abstract}

We present a search for near-infrared (3.5 - 5 $\mu$m) 
emission from baryonic dark matter in the form of low-mass stars
and/or brown dwarfs in the halo of the nearby edge-on spiral
galaxy NGC 5907. The observations were made using a 
256$\times$256 InSb array with a pixel scale of 17'' at
the focus of a liquid-helium-cooled telescope carried above the Earth's
atmosphere by a sounding rocket. In contrast to previous experiments
which have detected a halo around NGC 5907 in the V, R, I, J and K bands
at galactic radii 6 kpc $<$ r $<$ 10 kpc, our search finds no evidence for
emission from a halo at 10 kpc $<$ r $<$ 30 kpc. Assuming a halo
mass density scaling as r$^{-2}$, which is consistent with the flat rotation
curves that are observed out to radii of 32 kpc, the lower limit of the
mass-to-light ratio at 3.5-5 $\mu$m for the halo of NGC 5907 is 250 (2
$\sigma$) in solar units. This is comparable to the lower limit we have
found previously for NGC 4565. Based upon recent
models, our non-detection implies that hydrogen-burning stars contribute
less than 15$\%$ of the mass of the dark halo of NGC 5907.
Our results are consistent with the previous detection of extended
emission at r $\leq$ 10 kpc if the latter is caused by a stellar
population that has been ejected from the disk because of tidal
interactions. We conclude that the dark halo of NGC 5907, which is evident
from rotation curves that extend far beyond 10 kpc, is not made of
hydrogen burning stars.

\end{abstract}


\section{Introduction}

The observed flatness of rotation curves of spiral galaxies implies the
existence of massive halos of dark matter extending to r $\geq$ 30 kpc.
The rate of microlensing observed toward the LMC suggests that a
substantial fraction of the dark halo surrounding the Galaxy is in the
form of massive compact halo objects (MACHOs) with masses in the range of
0.05-1 M$_{\odot}$ (Alcock et al. 1997). The nature of these objects is
not known but, due to their gravitational effects and the difficulty in
their detection, they must have high mass-to-light ratios and could be
black holes, cool low-mass stars, white dwarfs or brown dwarfs.
Low-mass stars and brown dwarfs are both potentially detectable in
near-infrared bands. 

NGC 5907 is a nearby edge-on Sb galaxy located at (R.A, decl.) = (15h
15.9m, 56$^{\circ}$ 19') with an inclination angle of 87$^{\circ}$. Its
distance, from Fisher and Tully (1981), is taken to be 11.4 Mpc (for
H$_{o}$ = 75 km s$^{-1}$Mpc$^{-1}$), giving a linear scale of 3.3 kpc
arcmin$^{-1}$. Barnaby and Thronson (1992) observed it in the
near-infrared (NIR) and reported its H band (1.68 $\mu$m) bulge-to-disk
luminosity ratio to be low, 0.05. Based on their models (Barnaby and
Thronson 1994), estimates for the H-band bulge light of this galaxy at 5,
10, 20 kpc from its center are 24, 6, and 1 nW m$^{-2}$sr$^{-1}$,
respectively. Sackett et al. (1994) reported the detection in the R-band
of a faint halo of emission at r = 6 kpc about this galaxy. This result
was confirmed by Lequeux et al. (1996) in the V and I bands and by James
and Casali (1996) and Rudy et al. (1997) in the J and K bands. Rudy et
al. found the observed color to be peculiar and best modelled by low-mass
stars with near-solar metallicity. Recently, Shang et al. (1998) found a
faint ring extending to $\sim$40 kpc from the plane of the galaxy,
evidence for recent tidal disruption of the disk of NGC 5907, and raised
the issue of whether the emission detected at r$\sim$6 kpc is associated
with the dark halo, or with a disrupted component of the disk.

Most recently, Zheng et al.(1999) have obtained deep
intermediate-band photometry to distances of $\sim$ 15 kpc at 6660 \r{A}
and 8020 \r{A}, which can be compared to the wider bands R and I. Their
results are in good agreement with previous, published results, yet their
analysis concludes that any excess light seen does not originate from the
inferred symmetrical dark halo. Instead, they attribute the asymmetric
surface brightness as an artifact of the faint ring and of residual light
from imperfectly masked stars. Finally, Zepf et al. (2000) imaged a region
at 4.4 kpc from the galaxy center with the NICMOS instrument on the HST
and detected far fewer stars than expected for a normal stellar population
with the expected near-solar metallicity. The connection between the faint
extended emission around NGC 5907 and the halo of matter that extends to r
$\geq$ 30 kpc remains in doubt.

A spherical halo that produces a flat rotation curve is constrained to
have a mass density proportional to r$^{-2}$. Therefore, assuming that the
mass-to-light ration (M/L) of the halo is constant implies that the
surface brightness of the halo will scale as r$^{-1}$. Emission from the
disk and bulge of the galaxy drops off more rapidly with radius than the
required r$^{-1}$ of such a halo. Therefore any emission from the halo
should be separated more easily from that of the disk and bulge at large
galactic radii; the halo would be easier to detect against the galaxy.
Experiments sensitive to larger radii, (Uemizu et al.
(1998), Bock et al. (1998)), are also less likely to be contaminated by
components of the disk and allow for an accurate determination of the sky
brightness - important factors in light of recent work by Zheng et al.
(1999). Consequently, deep NIR observations of the galaxy extending to
larger radii are important to answer the question of the existence and
nature of NGC 5907's halo. Unfortunately, most ground-based observations
have limited sensitivity to a halo at large radii because of the
difficulty in detecting low surface brightness against the night-sky
background.

\section{Instrument}

NITE, the Near-Infrared Telescope Experiment, is a LHe-cooled,
wide-field, 3.5-5$\mu$m camera that is carried above the Earth's
atmosphere by a sounding rocket. The background against which a faint
halo must be detected is dominated by the zodiacal light which is at a
minimum at these wavelengths. This wavelength band is also near optimum
for detecting low-mass stars and brown dwarfs, both because they are
relatively brighter in this band and because the disk and bulge of the
galaxy are relatively dimmer. A halo formed of these objects may be
detected more easily at longer wavelengths than the optical bands used
from the ground. As well, recent models (Allard et al. 1997) suggest that
searches at 4.5-5$\mu$m should offer the best possibility for finding
brown dwarfs.

The optical system consists of a 16.5 cm, f/2.2-folded Gregorian
telescope. The 256$\times$256 indium-antimonide (InSb) array is
located at the focal plane. The image scale is 17'' pixel$^{-1}$;
the total field is 1.2$^{\circ}$$\times$1.2$^{\circ}$. The array is
continuously read out at a 3.8 Hz frame rate using a ``sampling up
the ramp'' technique implemented to simplify the flight
electronics and provide the maximum flexibility in post-flight data
analysis. The array is reset every 10s to avoid saturation on bright
stars. The entire optical system is cooled below 30 K within a
supercritical helium cryostat. The InSb array is maintained at 30 K to
optimize its performance. Previous results obtained with NITE are reported
in Uemizu et al. (1998). Further details concerning the instrument are
given by Bock et al. (1998).

\section{Observation}

The instrument was launched from the White Sands Missile Range,
New Mexico, at 00:22 (MDT), 1998 May 22 on a Black Brant IX
(36.175 UR) sounding rocket. The payload separated from the
rocket motor at 86 s after launch and the experiment door opened. The cooled
telescope extended outside of the payload at 90
s so that the cooled baffle tube would not view the emission from
the warm payload door. At 132 s the lid covering the telescope aperture
deployed, and the cold shutter in front of the field
stop opened, commencing observations of the sky. At
473 s, the shutter closed, terminating the observation.
The telescope observed NGC 5907 during most of this time. During the flight,
the detector temperature remained stable, varying less than 0.2$\%$ during
the observations. The temperatures of the cooled mirror and baffle tube were
also stable throughout the observations.

In order to accurately correct variations in responsivity across the array,
the galaxy was placed sequentially at four different positions across the
array, acquiring two 10 s exposures at each position. We intended to
image the galaxy at the center of each quadrant of the array, but an
offset in the attitude control system shifted the images toward one corner
of the array. After one cycle of eight exposures, the positions were
deliberately offset by several pixels for the second cycle, for a total of
sixteen 10 s exposures. We observed a contaminating source of emission,
probably caused by emission from the tumbling booster, during the middle
of the second image cycle. This contaminated the sky background
sufficiently (raising the background flux from 320 to 720 nW
m$^{-2}$sr$^{-1}$ from exposures 10 to 16, see Figure 1) that only the
first 10 exposures were used in the analysis. The bright calibration star
$\iota$ Dra, separated from the galaxy by $\sim$3.3 degrees, was observed
at the beginning and at the end of the observations. The first observation
of $\iota$ Dra, which was well-centered on the array, as well as the guide
star $\alpha$ Boo, were used to characterize the point-spread function and
to calibrate the camera.

\section{Data Reduction}

We processed the images to remove instrumental and environmental
effects. A \emph{master dark frame} was made from a set of 11 dark
frames obtained $\sim$30 minutes before launch with the cold shutter closed.
The value of each pixel in the master dark frame was determined by taking 
the median of the dark frame values at that pixel and iteratively removing 
outlying (more than 3$\sigma$ from the median) values (\emph{3$\sigma$
clipping}). We formed \emph{dark-subtracted frames} by subtracting this
master dark frame from each of the 10 frames used in the analysis. We
removed point sources from each of these dark-subtracted frames by iterative
3$\sigma$ clipping. The average flux in the remaining pixels was taken to be
the average \emph{sky level} of each of the frames. 

A relative gain correction (\emph{flat-field frame}) was derived from the
sky
data. Each of the dark-subtracted frames was normalized by its average sky
level. A large region (30 kpc) about the galaxy was masked out.
The average of the 10 frames at each position in the array was iteratively
3$\sigma$ clipped to produce a preliminary \emph{sky flat}.
Pixels where the 3$\sigma$ clipping left no remaining data for either the
master dark frame or the sky flat were not used in subsequent analysis.

Each dark-subtracted frame was divided by the preliminary sky flat and then
again iteratively 3$\sigma$ clipped to produce the flat-fielded sky
levels. These levels were removed from the dark-subtracted, flat-fielded
frames. A mosaic image was formed from these 10 frames. The frames were
registered relative to one another by iteratively offsetting images to
maximize the correlation function of the two images.

We measured the point-spread function (PSF) using a
combination of bright (but not saturated) field stars, the calibration star
$\iota$ Dra and the bright guide star $\alpha$ Boo. The PSF drops to
5$\times$10$^{-7}$ of its maximum at a radius of 28'. The PSF has a faint
extended envelope that is likely caused by imperfections on the primary
mirror. The PSF is approximately circular with a radial dependence of
r$^{-4.9}$ + 0.021r$^{-2.2}$ as shown in Figure 2.

A preliminary mask was generated by treating as foreground stars all pixels
$\ge$ 3 $\sigma$ above the sky background and masking out all pixels within 
a radius where the brightness of the point source would exceed 2.5 nW 
m$^{-2}$sr$^{-1}$ (0.5 of the sky background noise).

The process of creating the sky flat was iterated, applying the preliminary
mask at the outset in order to eliminate residual contamination in the sky
background levels and in the pixels used to create the sky flat. The new sky
flat had a small proportion ($\sim$6$\%$) of its pixels where no value could
be assigned due to a lack of data; for these pixels, there was no image that
did not contain a ``star''. These pixels were ignored in subsequent
analysis. The second sky flat and set of sky levels were used to produce the
final mosaic image, shown in Figure 3. The initial mosaic image had a
residual background level; the usefulness of reprocessing was evidenced by
the disappearance of the residual in the iterated image.

A final point-source mask was made from the iterated image in the manner
described above, and this mask was used to iterate the entire process of
forming the image one more time. Although the loss of data owing to a lack
of flat-field information increased in the second iteration, our analysis
of the
radial profile of the galaxy, described below, was stable. The result of the
first iteration, shown in Figure 3, was used in the final analysis.

To determine the surface brightness profile of the galaxy, as
described below, a $\pm$45$^{\circ}$ wedge was added to the final mask in
order to remove the contamination of the disk of NGC 5907 to the surface
brightness profile. The final image of the galaxy is shown with the mask
applied in Figure 4.

The surface brightness was calibrated to $\pm$ 3$\%$ (statistical
uncertainty) using photometry on the stars $\iota$ Dra (Johnson
1966) and $\alpha$ Boo (UKIRT standards, 
www.jach.hawaii.edu/ JACpublic/UKIRT/astronomy/calib/ukirt$\_$stds.html),
with zero-magnitude fluxes from Johnson (1965) and Cohen et al. (1992).
The PSF was used to correct for saturation. The 1 $\sigma$ dispersion of
the sky background per pixel in the regions of the combined image that we
use in this analysis is approximately 5 nW m$^{-2}$sr$^{-1}$. The measured
sky brightness is 291$\pm$10 nW m$^{-2}$sr$^{-1}$ in our band, which is in
rough agreement with DIRBE results at a solar elongation of 90$^{\circ}$
(Leinert et al. 1998).

\section{Results}

The surface brightness profile is produced by averaging pixels in circular
annuli, each 1 pixel (17'') wide, centered on the galaxy. The averaging
process used a set of relative weights based upon the number of individual
exposures used to obtain the pixel in the mosaiced image. The noise
calculated on the average used this set of relative weights as well. An
iterative 3$\sigma$ rejection on the pixels of each annulus eliminates any
residual point sources.

A spherically symmetric halo with a flat rotation curve has a mass density

\begin{displaymath}  \rho (r) =  \frac {V^{2}_{max}} 
{4 \pi Gr^2} \end{displaymath}

The mass column density distribution, $\Sigma$, is obtained by integrating
along a line of sight through the spherically symmetric halo.

\begin{displaymath} \Sigma (r) =  \frac {V^{2}_{max}}{4rG}
 \end{displaymath}

The rotation curve is measured out to a distance of 32 kpc; at large
radii the rotational velocity is 220  km s$^{-1}$ (Sancisi and van Alhada 
1987). Using this and the solar output in our band $\nu$L$_{\nu}$ = 
9.94$\times$10$^{24}$W implies:

\begin{displaymath} \nu I_{\nu} (r) = 705 \frac {1}{r(')
(\frac{M}{L})_{\odot}} nW m^{-2}sr^{-1}
 \end{displaymath}

The surface brightness of the galaxy should fall off as $\sim$
r$^{-1}$ at large radii if the emission traces the mass responsible
for the constant rotational velocity.

The measured flux from 1' to 35' (3.3-116 kpc) is fitted to a
four-parameter function, Ar$^{-n}$ + Br$^{-1}$ + C. The first term traces
the emission from the bulge and disk of the galaxy including the
effects of the finite PSF; the last term fits for any systematic residual
of the sky background after processing; and the second term, Br$^{-1}$,
tests for the existence of a halo following the mass distribution for a
flat rotation curve. Figure 5 plots the surface brightness profile of NGC
5907. The best fit function is overlaid on it. The best-fit values give A =
113.9$\pm$5.9 nW arcmin$^{3.21}$m$^{-2}$sr$^{-1}$, n = 3.21$\pm$0.13, B =
-0.37$\pm$1.38 nW arcmin m$^{-2}$sr$^{-1}$ and C = 0.009$\pm$0.071 nW
m$^{-2}$sr$^{-1}$. The reduced $\chi$$^{2}$ of the fit is 1.32 for 118
degrees of freedom. The result of $\chi$$^{2}$ greater than 1 may be due to
either residual structure from flat fielding errors at 1$\%$ of the sky
background, an imperfect model, or additional structure on the sky.

A possible source of bias for the halo component B is the wings of
the PSF. The extended wing is a positive component, and would tend to
force a false positive result for B. Therefore, a value of B consistent
with zero is not caused by the halo being hidden by the PSF. Our result is
a non-detection of the halo with a 1$\sigma$ statistical uncertainty of
1.4 nW arcmin m$^{-2}$sr$^{-1}$, corresponding to a brightness of $\sim$
0.1$\%$ of the zodiacal light background at a radius of $\sim$ 15 kpc.

\section{Discussion}

The term Br$^{-1}$ gives a non-detection of the luminosity density of the
halo with a 1$\sigma$ uncertainty of 1.4 nW arcmin m$^{-2}$sr$^{-1}$, and
a corresponding 2$\sigma$ lower limit of the mass-to-light (M/L) ratio of
the halo of NGC 5907 at large angular scales of 250 (M/L)$_{\odot}$ at the
photometric band of NITE. This result is very similar to the one obtained
from our observation in this band of NGC 4565, where we found the 2$\sigma$
lower limit of (M/L) to be 260 (M/L)$_{\odot}$ (Uemizu et al. 1998).

This limit on the M/L ratio of the halo at r $\sim$ 10 kpc lies well
above a simple extrapolation of the results derived from the halo detections 
at smaller radii and shorter wavelengths, as illustrated in Figure 6. 
Sackett et al. (1994), at R band, constrain 270 $\leq$ M/L$_{R,halo}$
$\leq$ 540. The results of Lequeux et al. (1996) report a color V-R=1.2
for the halo, resulting in 510 $\leq$ M/L$_{V,halo}$ $\leq$ 1000 and
V-I=1.35, resulting in 310 $\leq$ M/L$_{I,halo}$ $\leq$ 610.  Rudy et
al. (1997), report colors southwest of the galaxy as R-J=1.58, J-K=1.30
and northeast as
R-J=2.35, J-K=0.71. Averaging these colors results in 40 $\leq$
M/L$_{K,halo}$ $\leq$ 90 and 80 $\leq$ M/L$_{J,halo}$ $\leq$ 170.

Recent models and photometry of low-mass stars and brown dwarfs indicate the
objects have lower M/L at 4$\mu$m than in the optical or nearer IR. Saumon
et al. (1994) present calculations for infrared photometry of 
zero-metallicity low-mass stars (0.2-0.092 M$_{\odot}$). The recent
detection of the cool brown dwarf Gliese 229B (Gl 229B) by Nakajima et al.
(1995) has been reported and Leggett et al. (1999) gives revised photometry
at J, H, K, and L' bands. Figure 6 compares the measured halo M/L by Sackett
et al. (1994), Lequeux et al. (1996) and Rudy et al. (1997) as well as
Saumon et al. (1994)'s models and the data for Gl 229B and the brown dwarf
candidate GD 165B (Becklin \& Zuckerman 1988) with our limit for NGC
5907's halo.

Our lower limit on the M/L, taken alone, does not rule out a halo composed
entirely of objects similar to the brown dwarf Gl 229B. But when combined
with other results for the halo M/L the spectrum of mass-to-light ratios
including our limit is not consistent with or suggestive of any reasonable
low-mass dwarf candidate or model. In contast, Rudy et al. (1997) attempt to
model possible populations for the halo colors they measure with results
varying from low-mass stars to a significant accompanying giant population. 
The minimum hydrogen-burning star calculated by Saumon et al. (1994) has a
mass of 0.092 M$_{\odot}$ and M/L ratio at our band of 3.5-5$\mu$m is $\sim$
40. Therefore, our result indicates that low-mass stars at the
hydrogen-burning limit are responsible for less than $\sim$15$\%$ (at 2
$\sigma$) of the mass of the halo of NGC 5907.

If we assume that the detections of emission at shorter wavelengths and
smaller radii shown in Figure 6 are sampling the same population of sources,
then the spectrum of M/L for NGC 5907 does not present a consistent
picture for the halo. All of the positive detections have been made for r
$\le$ 10 kpc. At these radii, scattered light from the galaxy dominates our
radial profile through the effect of the PSF. We are sensitive at greater
radii (r $>$ 10 kpc), which previous observations do not probe, and where
the halo that produces the flat rotation curve (out to $\ge$ 30
kpc) is expected to dominate.

A possible resolution of the discrepancy is that the emission detected at
r$\leq$10 kpc by several groups is associated with a source that does not
dominate the halo at larger radii. Shang et al. (1998) measured a
significant warp along the galaxy to $\sim$40 kpc and an ellipse-shaped ring
structure perpendicular to the galaxy out to $\sim$40 kpc, suggestive of a
past interaction between NGC 5907 and a dwarf galaxy. A dwarf irregular
galaxy is present at the end of the observed warp in the disk of NGC 5907.
If this interaction has disrupted the bulge of the galaxy, the bulge's mass
density may drop off less steeply than the $\sim$r$^{-3}$ that is usually
observed. The edge of the bulge could then mimic a halo at small radii, but
may not extend throughout the whole of the dark halo. Therefore, the outer
halo could be composed of an entirely different population from the inner
halo, or nonbaryonic matter.

We find little evidence for interaction in our NIR images, as
would be expected since the interacting components should not be dominated
by NIR-bright low-mass stars. The data in Figure 4 in the 3 - 10
kpc range do not deviate from our fit by more than $\sim$100 in M/L. We do
not observe any features that correspond to the ellipse-shaped ring
reported by Shang et al. (1998).

Further uncertainty concerning the origin of the extended optical and
near-IR emission comes from Zepf et al. (2000), who seek to resolve the
emission into star counts in a deep HST NICMOS image 4.4 kpc from the
galaxy. They find far fewer stars than expected for a normal stellar
population with the high metallicity consistent with the red halo colors
reported by James \& Casali (1998) and by Lequeux et al. (1998), and
conclude that the halo must be metal-poor or dwarf-dominated, inconsistent
with an origin involving normal stellar populations, such as the
accretion of a metal-rich companion galaxy. Our non-detection casts further
doubt on the dwarf-dominated halo hypothesis.

Our result does not help resolve the debate regarding the origins of the
emission observed at radii $\leq$ 10 kpc. The previously reported halo
detections and colors were sensitive out to radii of 6-10 kpc. This could
be consistent with an interaction picture in which disparate inner and
outer halo regions exist. However, Zepf et al. (2000)'s result at 4.4 kpc
should be consistent with them. The analysis of Zheng et al. (1999)
demonstrates the difficulty in separating the halo light from contamination
by the ring feature and residual starlight imperfectly masked using the PSF.
This suggests there may be significant systematic effects not accounted for
in previous detections of extended emission near the galaxy. If the colors
are inaccurate, Liu et al. (1999) conclude the halo at 4.4 kpc could have
a normal, metal-poor stellar population, consistent with an interaction
picture and with our result. Alternatively, if previous detections of the
halo are in error, we may simply be probing a halo without a significant
stellar population. Unless some of the data are in error, there does not
appear to be a single, simple interpretation of the observations of NGC
5907's halo at radii $\leq$ 10 kpc.

\section{Conclusions}

The search for emission from the extended halo at larger galactic radii (
$\geq$ 10 kpc) is less prone to contamination by a tidally disrupted disk.
The observations reported here set a stringent lower limit M/L$\ge$250 on
the halo of NGC 5907. This result is quite consistent with our earlier
non-detection for NGC 4565 (Uemizu et al. 1998). Based on models and
observations of low-mass stars, we rule out a halo with more than
15$\%$ of its mass in hydrogen-burning stars.

We are grateful to G. G. Fazio, W. J. Forrest, and J. L. Pipher for
supplying the focal-plane array detector, S. D. Price for supplying the
cryogenic system, B. P. Crill for developing the translation mechanism, and
the Instrument Development Center of the School of Science, Nagoya
University. We thank the NASA Wallops Flight Facility payload team and White
Sands Missile Range staff involved in our 1998 flight for their dedicated
efforts. This work was supported by a grant from the Caltech President's
Fund, by NASA grant NAG54079, by Grant-in-Aid 06402002 for Scientific
Research from the Japanese Ministry of Education, Science, Sports, and
Culture, by the HayakawaSatio Fund, and by the US-Japan Research Cooperative
Program (JSPS-NSF).

\clearpage

\figcaption{Sky background levels during the observations of the galaxy.
The time is given in seconds from the first acquisition of the galaxy, at
t$+$195s from launch. Apogee occured at t$+$298s. The
contaminating emission's effect can be clearly seen.}

\figcaption{The radial dependence of the point-spread function (PSF),
measured using a combination of bright field stars and the calibration stars
$\iota$ Draconis and $\alpha$ Bootes. The PSF is approximately circular with
a radial dependence given by the fit (solid line) of 0.041(r$^{-4.9}$ +
0.021r$^{-2.2}$).}

\figcaption{Gray-scale mosaic image of NGC 5907 at 3.5-5 $\mu$m after
flat-fielding. A badly vignetted area that could not be flat-fielded well is
masked out. The full mosaiced image is 1.8$^{\circ}$ $\times$ 1.8$^{\circ}$.
The full range of the gray scale corresponds to $\pm$10\% of the sky
background. North is to the upper left and east is to the upper right.}

\figcaption{Same as Fig.3, but with stars and the disk of the galaxy
masked as described in the text.}

\figcaption{The surface brightness profile along the minor axis of NGC
5907. The brightness is measured by a weighted average of pixels in
circular annuli of width 0.94 kpc (1 pixel). For D = 11.4 Mpc, the linear
scale is 1 kpc = 0.30'. The best fit of Ar$^{-n}$ + Br$^{-1}$ + C to the
data for r $\ge$ 3.3 kpc (1') is given by the solid line. The observed sky
flux level is indicated. The expected emission from the halo is shown by
dashed lines for various M/L ratios. Error bars indicate $\pm$1$\sigma$ for
each annulus, with 2$\sigma$ upper limits (arrows) for data with negative
values. The data at large radii are binned for clarity.}

\figcaption{The mass-to-light ratio of our lower limit of NGC 5907 halo at
3.5-5 $\mu$m, compared with other detections of the halo of NGC 5907
(bars). The R is from Sackett et  al. (1994), V and I from Lequeux et al.
(1996) and J and K from Rudy et al. (1997).  Also presented are values
from modeled photometric brightnesses of 0.092, 0.1 and 0.2 M$_{\odot}$
zero-metallicity dwarfs by Saumon et al. (1994) (dotted lines).
Observations of brown dwarf Gl 229B (Leggett et al. 1999) and the brown
dwarf candidate GD 165B (Jones et al. 1994) are displayed (dashed lines),
with the mass of GD 165B assumed to be 0.08 M$_{\odot}$.}

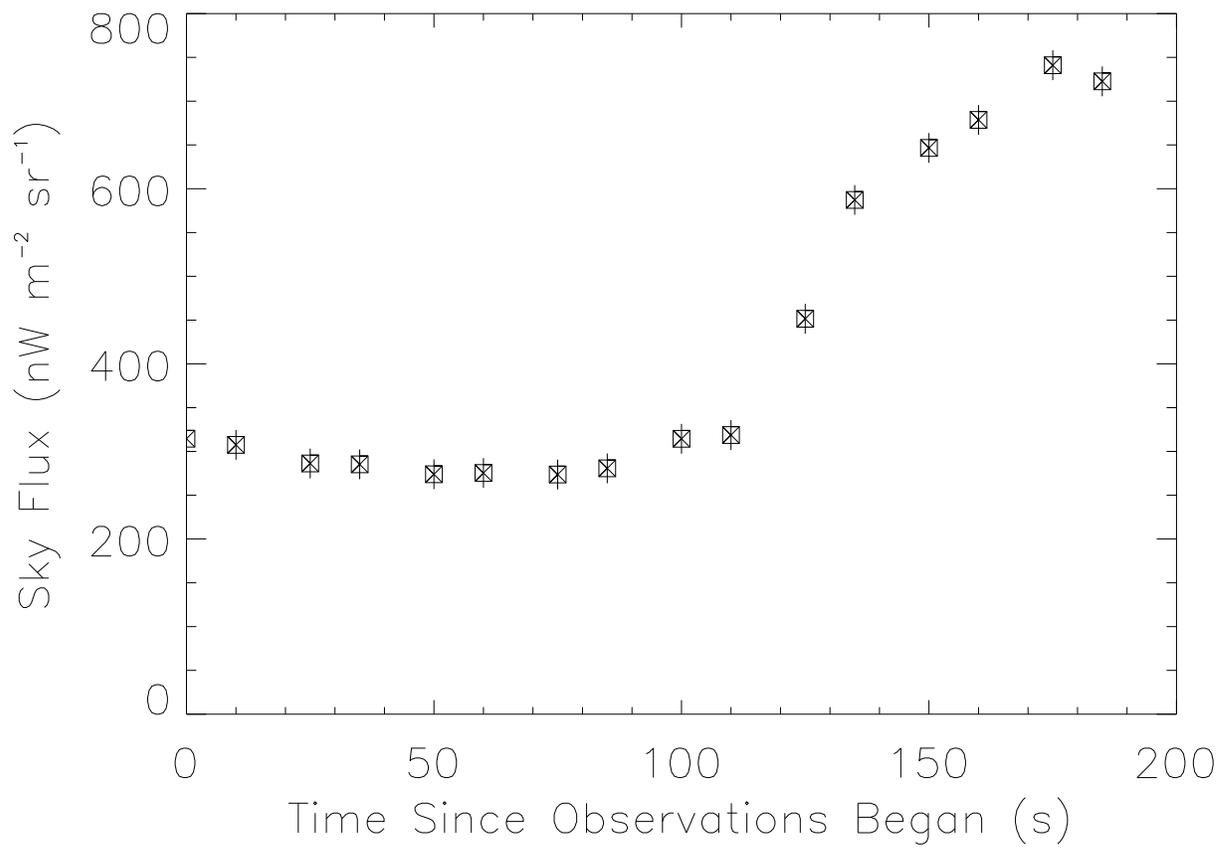
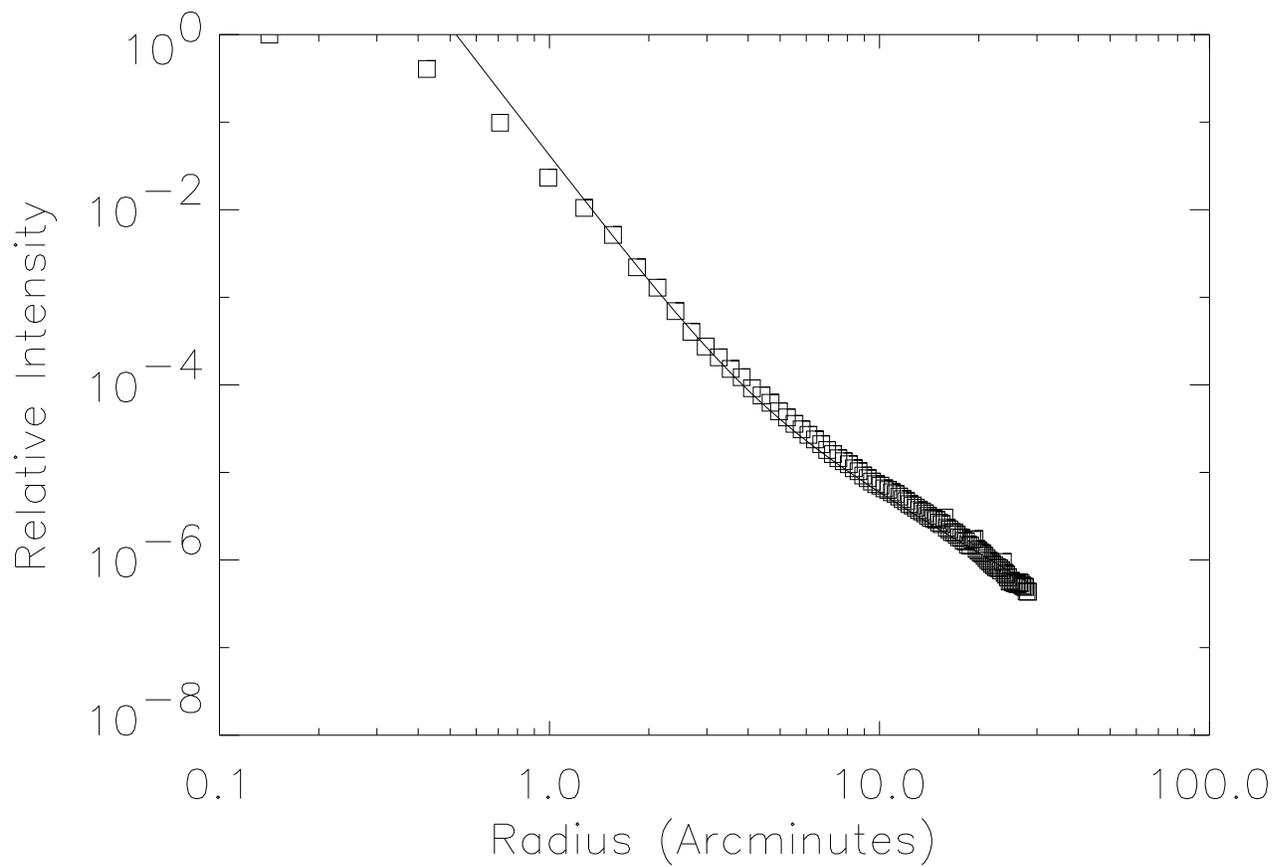
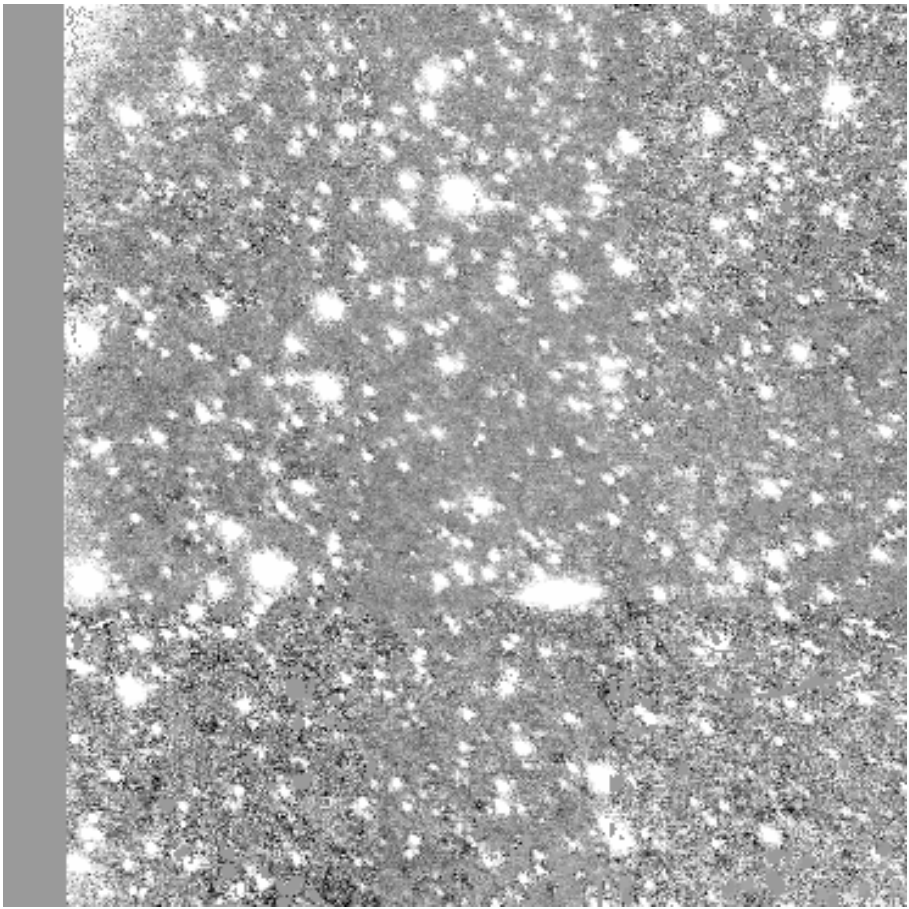
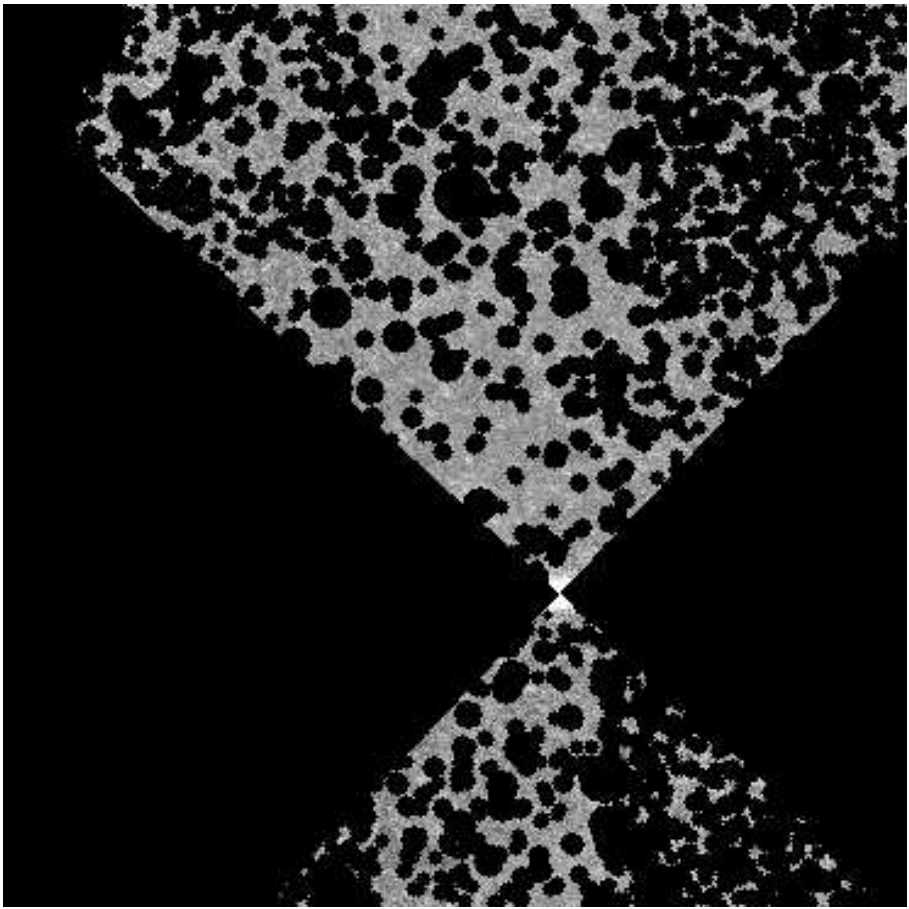
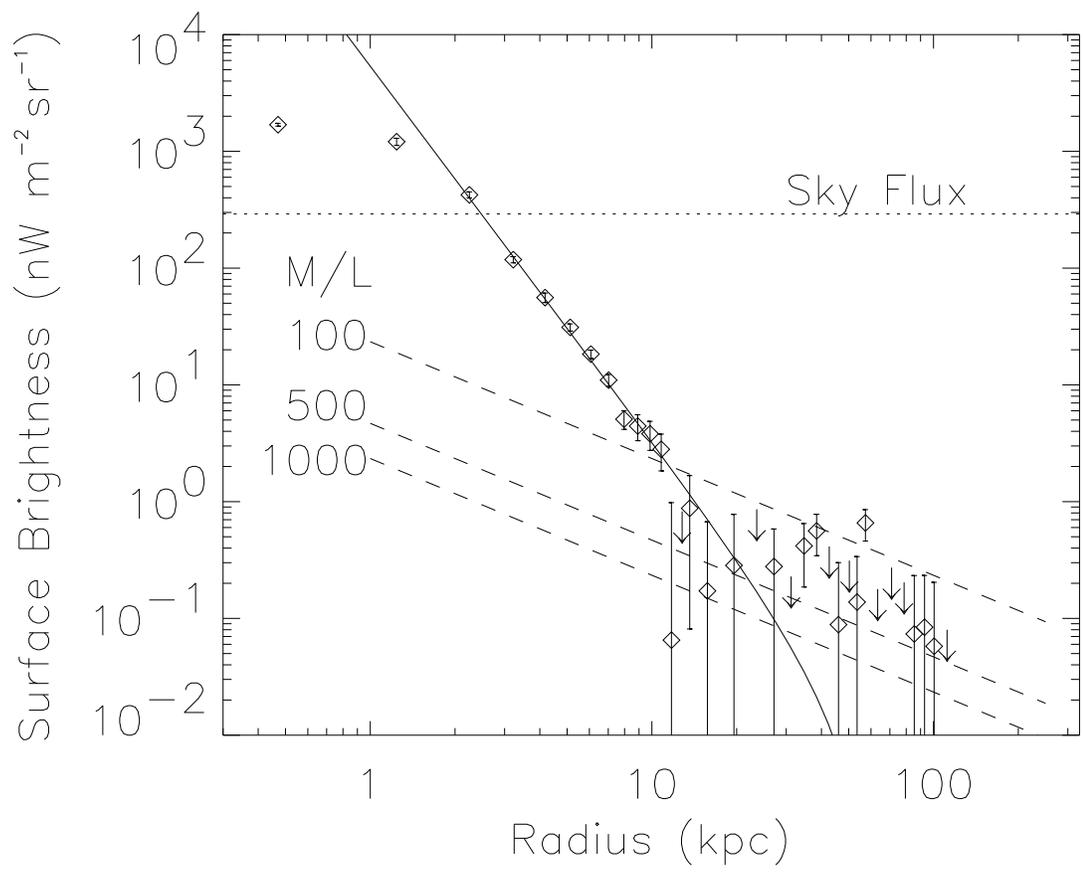
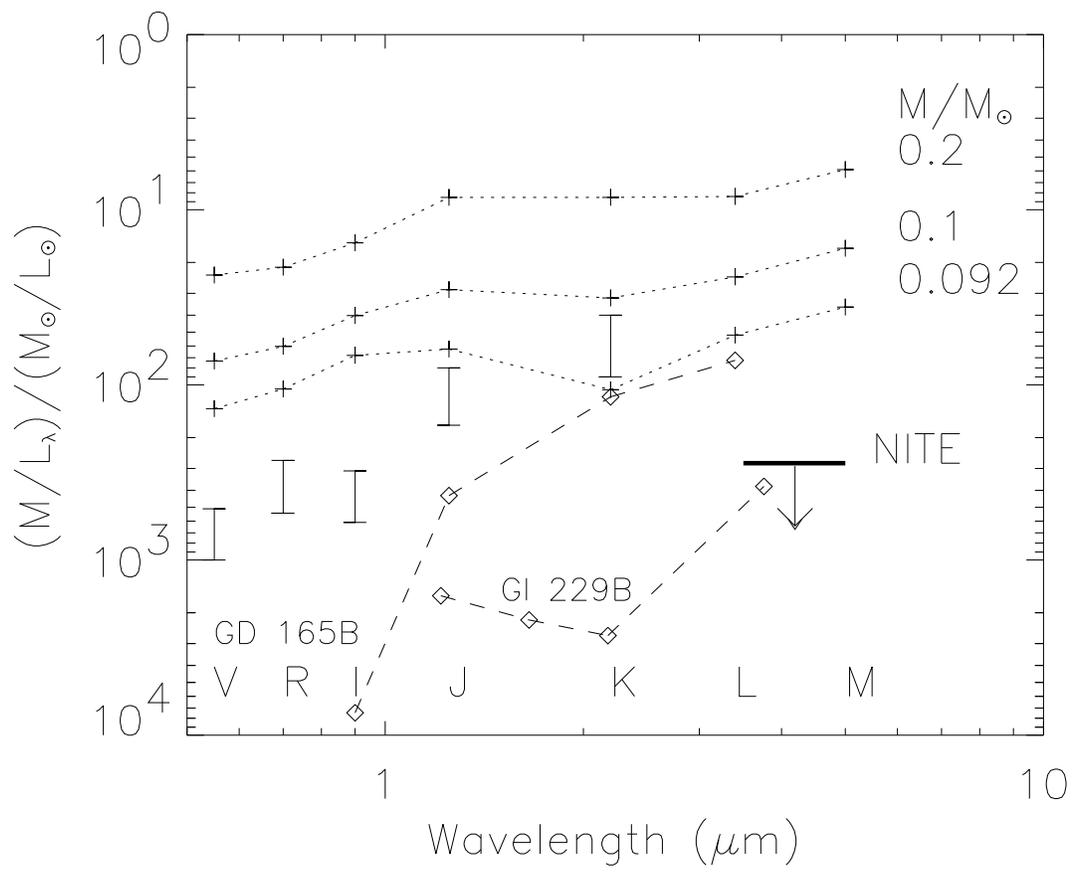
\end{document}